\documentclass[preprint,12pt]{elsarticle}
\usepackage{amsmath}
\usepackage{amsfonts}
\usepackage{bm}
\usepackage{lipsum}
\usepackage{enumerate}
\usepackage{graphicx}
\usepackage{amsfonts}

\usepackage{mathrsfs}
\usepackage{subfigure}
\usepackage{epstopdf}
\usepackage{url}
\usepackage{verbatim}
\usepackage{amssymb}
\usepackage{hyperref}
\usepackage{color}
\usepackage{listings}

\usepackage[T1]{fontenc}

\begin{document}

\begin{frontmatter}



\title{Persisting asymmetry in the probability distribution
function for a random advection-diffusion equation in impermeable channels}

\author[1]{Roberto Camassa}
\ead{camassa@amath.unc.edu}
\author[1]{Lingyun Ding}
\ead{dingly@live.unc.edu}
\author[2]{ Zeliha Kilic}
\ead{zkilic@asu.edu}
\author[1]{Richard M. McLaughlin \corref{mycorrespondingauthor}}
\cortext[mycorrespondingauthor]{Corresponding author}
\ead{rmm@email.unc.edu}
\address[1]{Department of Mathematics, University of North Carolina, Chapel Hill, NC, 27599, United States}
\address[2]{Center for Biological Physics, Arizona State University, Tempe, AZ, 85282, United States}





\begin{abstract}
In this paper, we study the effect of impermeable boundaries on the symmetry properties of a random passive scalar field advected by random flows. We focus on a broad class of nonlinear shear flows multiplied by a stationary, Ornstein-Uhlenbeck (OU) time varying process, including some of their limiting cases, such as Gaussian white noise or plug flows.  
For the former case with linear shear, recent studies \cite{camassa2019symmetry} numerically demonstrated that the decaying passive scalar's long time limiting probability distribution function (PDF) could be negatively skewed in the presence of impermeable channel boundaries, in contrast to rigorous results in free space which established the limiting PDF is positively skewed \cite{mclaughlin1996explicit}.  Here, the role of boundaries in setting the long time limiting skewness of the PDF is established rigorously for the above class using the long time asymptotic expansion of the $N$-point correlator of the random field obtained from the ground state eigenvalue perturbation approach proposed in \cite{bronski1997scalar}.  Our analytical result verifies the conclusion for the linear shear flow obtained from numerical simulations in \cite{camassa2019symmetry}. Moreover, we demonstrate that the limiting distribution is negatively skewed for any shear flow at sufficiently low P\'eclet number.  We demonstrate the convergence of the Ornstein-Uhlenbeck case to the white noise case in the limit $\gamma\rightarrow \infty$ of the OU damping parameter, which generalizes to the channel domain problem the results for  free space in \cite{resnick1996dynamical}. We show that the long time limit of the first three moments depends explicitly on the value of $\gamma$, which is in contrast to the conclusion in \cite{vanden2001non} for the limiting PDF in free space.  To find a benchmark for theoretical analysis, we derive the exact formula of the $N$-point correlator for a flow with no spatial dependence and Gaussian temporal fluctuation, generalizing the results of \cite{bronski2007explicit}. The long time analysis of this formula is consistent with our theory for a general shear flow. All results are verified by Monte-Carlo simulations.

\end{abstract}




\end{frontmatter}

\section{Introduction}
Partial differential equations (PDEs) with random coefficients have been the focus of many studies as they occur in a variety of mathematical models of physical systems. 
Some examples from this class of PDEs include passive scalar  (e.g., fluid temperature or solute concentration) advection by random fluid flows~\cite{mclaughlin1996explicit,balkovsky1998two,chertkov1998intermittent,sinai1989limiting}, linear and nonlinear Schr\"{o}dinger equations with random
potentials~\cite{anderson1958absence,bronski1997stability}, light propagating through random media~\cite{stephen1988temporal} and random water waves impinging on a step \cite{bolles2019anomalous,majda2019statistical}.

Motivated by the Chicago convection experiments \cite{castaing1989scaling}, random passive scalars have been intensely studied as a simplified model for intermittency in fluid turbulence that, while enjoying a linear evolution, retain many statistical closure features reminiscent of problems in fluid turbulence \cite{sinai1989limiting,yakhot1990phenomenological,pumir1991exponential,kerstein1991linear,kimura1993statistics,ching1994passive}.  
In particular, the case of diffusing passive scalar advected by a rapidly fluctuating Gaussian random fluid flow has 
been the focus of much analysis as the moment closure problem is bypassed in the white noise limit \cite{majda1993random,majda1993explicit,balkovsky1998two,bronski1997scalar,bronski2000rigorous,kraichnan1968small,majda1993random,mclaughlin1996explicit,sinai1989limiting}.  Notably, the availability of closed evolution equations for the statistical correlators led to the discovery that a diffusing passive scalar could inherit a heavy-tailed, non-Gaussian PDF from a Gaussian random fluid flow \cite{majda1993explicit,mclaughlin1996explicit,bronski2000rigorous,bronski2000problem}.   Additional studies have explored the role played by finite or infinite correlation times in a random shear flow \cite{resnick1996dynamical,vanden2001non}. This generic non-Gaussian behavior in a passive scalar has been termed `scalar intermittency'.  Subsequently, similar findings have been observed in field experimental data, such atmospheric wind measurements \cite{antonia1977log} as well as observations of stratospheric inert tracers~\cite{sparling2001scale}.

Further investigations have provided more in depth understanding of how the non-Gaussian measure is dynamically attained \cite{camassa2008evolution}, and further explored the case of a passive scalar advected by a shear-free temporally fluctuating wind, where the entire probability measure can be determined at any time \cite{bronski2007explicit}. This further exhibited how the diffusivity adjusts the location of singularities in the probability measure.  Additional studies contrasted the scalar PDF inherited by an unbounded linear shear with that of a bounded, periodic shear flow \cite{bronski1997scalar}.  This established that for integrable random initial data the PDF would `Gaussianize' at long times, whereas short ranged, random wave initial data would produce divergent flatness factors in the same limit.  

While theoretically interesting, unbounded domains are of course unattainable in actual experiments, and the effects of boundaries need to be included for realistic models. Recently, the role of impermeable boundaries has been explored in a channel geometry with deterministic initial conditions \cite{camassa2019symmetry}.   This work established the surprising role that the boundary conditions play in setting the skewness of the PDF.  McLaughlin and Majda \cite{mclaughlin1996explicit} established that in free-space, with deterministic initial data, the long time PDF skewness would be strictly positive, whereas Monte-Carlo simulations in \cite{camassa2019symmetry} have demonstrated that with no-flux boundary conditions in a channel geometry, the long time PDF skewness can be negative.  Further, it has been shown in \cite{camassa2019symmetry} that such flows could be physically realized by a randomly moving wall.  More recently, the enhanced diffusion \cite{taylor1953dispersion} and third spatial Aris moment \cite{aris1956dispersion} induced by a periodically moving wall was studied experimentally and theoretically \cite{ding2020enhanced}, where it is noteworthy that the flows' temporal dependence is non-multiplicative.   

Inspired from the ground state energy expansion strategy to handle more realistic flows (e.g. periodic flows in~\cite{bronski1997scalar}) and the recent numerical findings provided in~\cite{camassa2019symmetry}, here we rigorously establish that impermeable boundary conditions in a channel geometry can yield a scalar PDF with negative long-time skewness. We do so for a range of molecular diffusivities and for arbitrary nonlinear shear layers multiplied by a stationary Ornstein-Uhlenbeck process, through an explicit calculation of the long time scalar skewness asymptotics.  Further, we gain insight into the role of the correlation time in the underlying stochastic process in the dynamic evolution of the scalar skewness, and in particular establish that longer correlations times yield increased transient dynamics.  

The paper is organized as follows: In section \ref{sec:setup}, we formulate the problem of the evolution of the passive scalar field advected by a nonlinear shear layers multiplied by an Ornstein-Uhlenbeck random process with an impermeable boundary and introduce some important conclusions of this scalar intermittency model. In section \ref{sec:GroundStateEnergyExpansion}, we derive a long time asymptotic expansion of the $N$-point correlation function of the scalar field by the perturbation theory and the differential operator spectral theory. Based on the $N$-point correlation function, we study the PDF of the scalar and show how the flow controls the asymmetry of PDF, which rigorizes and generalizes the conclusions in the article \cite{camassa2019symmetry}. In section \ref{sec:wind}, we study the model with a spatially uniform, temporally Gaussian fluctuated shear flow. Being a special case of shear flow, the spatially uniform structure allows access to the exact formulae of the Green's function and the $N$-point correlation function. These are consistent with the long time asymptotic expansion in section \ref{sec:setup} and are verified by Direct Monte-Carlo (DMC) simulation proposed in \cite{camassa2019symmetry}.  In section \ref{sec:numerical}, we perform numerical simulations for spatially non-uniform flows using the backward Monte-Carlo method. The numerical results quantitatively demonstrate the validity of the formulae we derive in section \ref{sec:setup}.  In section \ref{sec:discuss}, we summarize the conclusions from the findings in the paper and  briefly discuss future studies.

\section{Setup and background of the problem for scalar intermittency}
\label{sec:setup}

We will study intermittency in the following non-dimensional, random advection diffusion equation with deterministic initial condition $T_{0}\left(x,y\right)$ and impermeable channel
boundary conditions,
\begin{equation}\label{eq:advectionDiffusion}
\displaystyle \frac{\partial T}{\partial t}+\xi(t)u(y)\frac{\partial T}{\partial x}=\kappa \Delta T \,,\qquad
 T(x,y,0)=T_{0}(x,y)\,, \qquad
\displaystyle\left. \frac{\partial T}{\partial y}\right|_{y= 0,L}=0 \, , 
\end{equation}
where the domain is $\left\{ (x,y)| x\in \mathbb{R}, y \in  [0, L] \right\}$, $L$ is the gap thickness of the channel, $\kappa$ is the diffusivity, $\xi(t)$ is a  zero-mean, Gaussian random process with the correlation function given by $\left\langle \xi(t)\xi(s) \right\rangle=R(t,s)$.
The time dependent random shear flow  can originate from either a time varying pressure field, or by randomly moving portions of the boundary. Such a shear flow can be obtained by solving the Navier-Stokes equations with boundary conditions matching the wall velocity $\xi(t)$, see section 2 of \cite{camassa2019symmetry} for more details.  We note that in this study we only consider shear flows whose spatial averages are non-zero, such as would arise in an experiment in which only one channel wall is moved, with statistics measured in the laboratory frame.  We note that the symmetric case involving two oppositely moving walls requires higher order asymptotics to compute leading order long time skewness limits and will be explored in future work. 
In this paper, two additional simplifying assumptions are made. 1) $\xi(t)$ is a Gaussian white noise in time so that $R(t,s)=g \delta(t-s)$, or 2) $\xi(t)$ is a stationary Ornstein-Uhlenbeck process with damping $\gamma$ and dispersion $\sigma$, which is the solution of stochastic differential equation (SDE) $\mathrm{d}\xi (t) =-\gamma \xi (t)\mathrm{d}t +\sigma \mathrm{d}B (t)$ with initial condition $\xi(0) \sim \mathcal{N} (0, {\sigma^2}/{2 \gamma})$. Here $B (t)$ is the standard Brownian motion and $\mathcal{N} (a,b)$ is the normal distribution with mean $a$ and variance $b$.   The correlation function of $\xi (t)$ is $R(t,s)=\frac{\sigma^{2}}{2\gamma}e^{-\gamma \left| t-s \right|}$. $\gamma^{-1}$ is often referred to as the correlation time of the OU process. It is easy to check that the stationary Ornstein-Uhlenbeck process converges to the Gaussian white noise process as the correlation time vanishes with fixed ${\sigma}/{\gamma}$.   

Notice that $\gamma \sim \frac{1}{\text{Time}}$, $\sigma \sim \frac{1}{\text{Time}^{ \frac{1}{2}}}$. With the change of variables, 
\begin{equation}
\begin{array}{ccc}
Lx'= x \quad Ly'=y&\frac{L^2}{\kappa}t'=t & g=\frac{\sigma}{\gamma} \\
\frac{\kappa}{L^2}\gamma'=\gamma  &U= \frac{L}{g^{2}} &v (y,t)= u (y)\xi (t)  \\
Uv'(y,t)=v (y, t) &\frac{g \sqrt{\kappa}}{L}\xi'(\frac{L^2}{\kappa}t')=\xi (t)  &T' (Lx',Ly',\frac{L^2}{\kappa}t')=T (x,y,t) \\
\end{array}
\end{equation}
We can drop the primes without confusion and obtain the nondimensionalized version of \eqref{eq:advectionDiffusion}:
\begin{equation}\label{eq:advectionDiffusionNonDimension}
\displaystyle \frac{\partial T}{\partial t}+\text{Pe} \,\xi(t)u(y)\frac{\partial T}{\partial x}=\Delta T\,,\qquad
\displaystyle T(x,y,0)=T_{0}(x,y)\,,\qquad
\displaystyle \left.\frac{\partial T}{\partial x}\right|_{x= 0,1}=0\,,
\end{equation}
where the domain is $\left\{ (x,y)| x\in \mathbb{R}, y \in  [0,1] \right\}$, and we have introduced the P\'{e}clet number  $\text{Pe}=  {U L}/{ \kappa}=  {L^{2}}/{ (g^{2}\kappa)}$. When $\xi (t)$ is the white noise process, the correlation function of $\xi (t)$ is $R(t,s)= \delta (t-s)$.  Conversely, when $\xi (t)$ is the stationary Ornstein-Uhlenbeck process, the underlying SDE becomes $\mathrm{d}\xi (t) =-\gamma \xi (t)\mathrm{d}t + \mathrm{d}B (t)$ with the initial condition $\xi (0) \sim \mathcal{N} (0,\frac{\gamma}{2})$, and the correlation function of $\xi (t)$ is $R(t,s)= \frac{\gamma}{2}e^{-\gamma \left| t-s \right|}$.

Define the $N$-point correlation function $\mathbf{\Psi}_{N}$ of the scalar field $T(x,y,t)$: $\mathbb{R}^{2N}\times \mathbb{R}^{+}\rightarrow \mathbb{R}$ by $\mathbf{\Psi}_N (\mathbf{x}, \mathbf{y},t) =\left<\prod_{j=1}^N T(x_j,y_j,t)\right >_{\xi (t)}$, where $\mathbf{x}=\left(x_1,x_2,\cdots,x_N\right)$, $\mathbf{y}=(y_1,y_2,\cdots,y_N)$. Here, the brackets $\left\langle \cdot  \right\rangle_{\xi(t)}$ denote ensemble averaging with respect to the stochastic process $\xi(t)$.  The $\mathbf{\Psi}_{N}$ associated with the free space version of \eqref{eq:advectionDiffusion} is known for some special flows. When $\xi(t)$ is  the Gaussian white noise process, Majda \cite{majda1993random} showed that $\hat{\mathbf{\Psi}}_{N}$ satisfies a $N$-body parabolic quantum mechanics problem,
\begin{eqnarray}\label{closureeqnWhite}
\frac{\partial \hat{\mathbf{\Psi}}_{N}}{\partial t} &=& \Delta_N \hat{\mathbf{\Psi}}_{N}- \left( \frac{\text{Pe}^2}{2}\nonumber \left( \sum\limits_{j=1}^{N}u\left(y_{j}\right) k_{j}\right)^2+ |\mathbf{k}|^2  \right)\hat{\mathbf{\Psi}}_{N}\\
\nonumber \hat{\mathbf{\Psi}}_{N}(\mathbf{k},\mathbf{y},0)&=& \prod_{j=1}^N \hat{T}_{0}(k_j,y_j)\\
\end{eqnarray}
where $\hat{f}(\mathbf{k})= \int\limits_{\mathbb{R}^N}^{} \mathrm{d}\mathbf{y} e^{\mathrm{i}(\mathbf{x}\cdot \mathbf{k} )}f(\mathbf{x})$ is the Fourier transformation of $f(\mathbf{x})$, $\Delta_N$ is the Laplacian operator in $N$ dimensions $\Delta_{N}=\sum\limits_{j=1}^N \frac{\partial^2}{\partial y_j^{2}}$,  $\mathbf{k}=\left(k_1,k_2,\cdots,k_N\right)$. When $u (y)=y$, Majda \cite{majda1993random}  derived the exact expression of $\Psi_N$.  A rotation of coordinates reduces the $N$-dimensional problem to a one-dimensional problem. Then the solution of \eqref{closureeqnWhite} is available via Mehler's formula.
Based on this exact $N$-point correlation function, the distribution of the scalar field advected by a linear shear flow has been studied for deterministic and random initial data. The non-Gaussian behaviors of PDF have been reported in  \cite{mclaughlin1996explicit,bronski2000problem,bronski2000rigorous}.

When $\xi(t)$ is the stationary Ornstein-Uhlenbeck process, by introducing an extra variable $z$, Resnick \cite{resnick1996dynamical} showed  that $\hat{\mathbf{\Psi}}_{N} (\mathbf{k}, \mathbf{y},t)= \frac{1}{\sqrt{\pi}}\int\limits_{-\infty}^{+\infty} \mathrm{d} z \hat{\psi} (\mathbf{k}, \mathbf{y},z,t) e^{-z^2}$, where $\hat{\psi} (\mathbf{k}, \mathbf{y},z,t)$ satisfies the following partial differential equation:
\begin{eqnarray}
\frac{\partial \hat{\psi}}{\partial t}+\mathrm{i} \text{Pe}\sqrt{\gamma} z \sum\limits_{j=1}^N k_{i} u(y_{i}) \hat{\psi}+ \gamma z \hat{\psi}_z &=& \Delta_N \hat{\psi}- |\mathbf{k}|^{2}\hat{\psi}+ \frac{\gamma}{2}\hat{\psi}_{zz}\\
\nonumber \hat{\psi}(\mathbf{k},\mathbf{y},z,0)&=& \prod_{j=1}^N \hat{T}_{0}(k_j,y_j)
\label{closureeqnOU}
\end{eqnarray}
When $u (y)=y$, Resnick derived the exact expression for $\Psi_{N}$ via the same strategy Majda used for solving \eqref{closureeqnWhite} and showed it converges to the solution of \eqref{closureeqnWhite} in the limit  $\gamma\rightarrow \infty$ of the damping OU parameter.

These results are all derived in free-space.   The $N$-point correlation function $\Psi_{N}$ for the boundary value problem \eqref{eq:advectionDiffusionNonDimension} is unknown even for simple-geometry domains.  For periodic boundary conditions, Bronski and McLaughlin \cite{bronski1997scalar} carried out a second order perturbation expansion for the ground state of periodic Schr\"odinger equations to analyze the inherited probability measure for a passive scalar field advected by periodic shear flows with multiplicative white noise.
In  \cite{camassa2019symmetry} equation \eqref{eq:advectionDiffusionNonDimension} was studied with the flow $u(y)=y\xi(t)$  where $\xi(t)$ is the white noise process. A dramatically different long time state resulting from the existence of the impermeable boundaries was found. In particular, the PDF of the scalar in the channel case has negative skewness,  in stark contrast to free space, where the limiting skewness is positive. Inspired by the observation reported in \cite{camassa2019symmetry}, we further explore here the PDF of the advected scalar in the presence of impermeable boundaries by the perturbation method introduced in \cite{bronski1997scalar}.  Briefly, the long time behavior of the Fourier transformation of $N$-point correlation function $\hat{\Psi}_{N}$ of the scalar field is dominated by the neighborhood of the zero frequency  $\mathbf{k}=\mathbf{0}$. This observation reduces the series expansion of $\hat{\Psi}_{N}$ to a single multi-dimensional Laplace type integral. Then, the standard asymptotic analysis and inverse Fourier transformation yield the long time asymptotic expansion of $\Psi_N$.


\section{Long-time asymptotics: ground state energy expansion in channel geometry}\label{sec:GroundStateEnergyExpansion}
For bounded domains, the $N$-point correlation function $\mathbf{\Psi}_{N}$  inherits the impermeable boundary condition from the scalar field.
From spectral theory of parabolic differential operators, the solution of \eqref{closureeqnWhite},\eqref{closureeqnOU} can be written as an eigenfunction expansion of the form
\begin{equation}
\begin{array}{rl}
\label{eq:eigenfunctionExpansion}
  \hat{\Psi}_{N}(\mathbf{k},\mathbf{y},t) =& \sum\limits_{l=0}^{\infty}\beta_l(\mathbf{k}) \phi_l (\mathbf{k},\mathbf{y})
          e^{-\lambda_l (\mathbf{k}) t} \\ 
\end{array}.
\end{equation}
When the statistics of velocity field is white in time, $\lambda_l, \phi_l$ are the eigenvalues and eigenfunctions of the eigenvalue problem
\begin{equation}
\begin{array}{rl}
-\lambda_{l}\phi_{l}&= \Delta_N \phi_{l}- \left( \frac{\text{Pe}^2}{2} \left( \sum\limits_{j=1}^{N}u\left(y_{j}\right) k_{j}\right)^2+ |\mathbf{k}|^2  \right)\phi_{l} ,\\
\frac{\partial \phi_{l}}{\partial y_{j}}|_{y_{j}= 0,1}&=0, \qquad \forall \,  1\leq j\leq N.
\end{array}
\end{equation}
For simplicity, we scale $\phi_l$ so that $\left\{ \phi_{l} \right\}_{l=0}^{\infty}$ form an orthonormal basis with respect to the inner product $\left\langle f(\mathbf{y}),g(\mathbf{y}) \right\rangle= \int\limits_{[0,1]^{N}}^{}\mathrm{d} \mathbf{y} f(\mathbf{y})g(\mathbf{y})$ for all $\mathbf{k}$. $\beta_l$ are determined by the initial condition and the eigenfunction via $\beta_l(\mathbf{k})= \left\langle \prod_{j=1}^N \hat{T}_{0}(k_j,y_j),  \phi_l(\mathbf{k},\mathbf{y}) \right\rangle  $.

When $\xi(t)$ is the stationary Ornstein-Uhlenbeck process, $\phi_l(\mathbf{k}, \mathbf{y}) = \frac{1}{\sqrt{\pi}} \int\limits_{-\infty}^{+\infty}\mathrm{d} z \varphi_{l} (\mathbf{k}, \mathbf{y},z) e^{-z^2}$, where $\lambda_l, \varphi_l$ are the eigenvalues and eigenfunctions of the eigenvalue problem
\begin{equation}
\begin{array}{rl}
-\lambda_{l} \varphi_{l}&=-\mathrm{i} \text{Pe}\sqrt{\gamma} z \sum\limits_{j=1}^N k_{i} u(y_{i}) \varphi_{l}- \gamma z \frac{\partial \varphi_{l}}{\partial z} + \Delta_N \varphi_{l}- |\mathbf{k}|^{2}\varphi_{l}+ \frac{\gamma}{2} \frac{\partial^2 \varphi_l}{\partial z^{2}} \\
\frac{\partial \varphi_{l}}{\partial y_{j}}|_{y_{j}= 0,1}&=0, \qquad \forall  1\leq j\leq N.
\end{array}
\end{equation}
We also choose $\varphi_{l}$ such that $\left\{ \varphi_{l} \right\}_{l=0}^{\infty}$ form an orthonormal basis with respect to the inner product $ \left\langle f(\mathbf{y},z),g(\mathbf{y},z) \right\rangle=\frac{1}{\sqrt{\pi}} \int\limits_{-\infty}^{+\infty}\mathrm{d}z \int\limits_{[0,1]^{N}}^{}\mathrm{d} \mathbf{y} f(\mathbf{y},z)g^{*}(\mathbf{y},z)e^{-z^{2}} $ respectively, where $g^{*}$ is the complex conjugate of $g$. $\beta_l$ have the same definition as the Gaussian white noise case.

Bronski and McLaughlin \cite{bronski1997scalar} proved that $\lambda_l(\mathbf{k})$ strictly increases with respect to the subscript $l$ for all $\mathbf{k}$ and have a global minimum value at $\mathbf{k}=\mathbf{0}$; in particular, $\lambda_{0}(\mathbf{0})=0, \lambda_{1}(\mathbf{0})=\pi^{2}$. As a consequence, the series given in (\ref{eq:eigenfunctionExpansion}) is dominated at long times by the ground state $j=0$, since the other terms
are $\mathcal{O} (e^{-\pi^{2}t})$. This observation yields the following asymptotic formula valid at long times for arbitrary $N$-point correlation function of the scalar field:
\begin{equation}
\begin{array}{rl}
\mathbf{\Psi}_{N}(\mathbf{x}, \mathbf{y},t)= \frac{1}{(2\pi)^{N}} \int\limits_{\mathbb{R}}^{} \mathrm{d}\mathbf{k} e^{-\mathrm{i}(\mathbf{x}\cdot \mathbf{k})}\beta_0(\mathbf{k}) \phi_0 (\mathbf{k},\mathbf{y})
          e^{-\lambda_0 (\mathbf{k}) t} +\mathcal{O} (e^{-\pi^{2}t}) \text{ as } t\rightarrow \infty.
\end{array}
\end{equation}
This is an $N$-dimensional Laplace type integral with respect to the frequency variable $\mathbf{k}$. By formula (1) in \cite{inglot2014simple}, the asymptotic expansion of $\mathbf{\Psi}_N(\mathbf{x}, \mathbf{y},t)$ for large $t$ is
\begin{equation}
\begin{array}{rl}
\label{eq:NpointCorrelation}
\mathbf{\Psi}_N (\mathbf{x}, \mathbf{y},t)= & \frac{1}{\left( 2\pi t \right)^{\frac{N}{2}} \text{det} (\mathbf{H})} \left(\int\limits_{0}^{1}\mathrm{d} y\hat{T}_{0} (0,y) \right)^{N} +\mathcal{O}(t^{\frac{N+2}{2}}) \text{ as } t\rightarrow \infty, \\
\end{array}
\end{equation}
where $\mathbf{H}_{i,j}= \frac{\partial^2 }{\partial k_i \partial k_j} \lambda_0(\mathbf{k},\mathbf{y})|_{\mathbf{k}=\mathbf{0}}$ is the Hessian matrix of the eigenvalue $\lambda_0 (\mathbf{k})$ at $\mathbf{k}=\mathbf{0}$.

Here, we are primarily concerned with single-point statistics, namely the moment of the random scalar field at point $(x,y)$, $\left\langle T^{N}(x,y,t) \right\rangle=\mathbf{\Psi}_N (\mathbf{x},\mathbf{y},t)$ , where all components of $\mathbf{x}, \mathbf{y}$ are $x,y$, namely $x=x_{1}=x_{2}=...=x_{N}, y=y_{1}=y_{2}=...=y_{N}$. When $N=1,2,3$, the $\text{det} (\mathbf{H})$ in the \eqref{eq:NpointCorrelation} only depend on the derivative of eigenvalues in the one-dimensional eigenvalue problem $\lambda^{(2)}=\frac{\partial^{2}}{\partial k_1^{2}}\lambda_0 (k_{1})|_{k_1=0} $ and the derivative of eigenvalues in the two-dimensional eigenvalue problem $\lambda^{(1,1)}=\frac{\partial^{2}}{\partial k_1\partial k_2}\lambda_0 (k_{1},k_2)|_{k_1=0,k_2=0} $. Hence, as $t\rightarrow \infty$, the first three moments are
\begin{equation}\label{eq:firstThreeMoments}
\begin{array}{rl}
\left\langle T (x,y,t) \right\rangle =&\int\limits_{0}^{1}\mathrm{d} y\hat{T}_{0} (0,y)  
\displaystyle \frac{1}{(2\pi t)^{\frac{1}{2}}\sqrt{\lambda^{(2)}}}+\mathcal{O} (t^{- \frac{3}{2}}) \\
\left\langle T^{2}(x,y,t) \right\rangle =& \left( \int\limits_{0}^{1}\mathrm{d} y\hat{T}_{0} (0,y) \right)^{2} 
\displaystyle \frac{1}{2\pi t  \sqrt{(\lambda^{(2)})^{2}-(\lambda^{(1,1)})^{2} }}+\mathcal{O} (t^{-2}) \\
\left\langle T^{3}(x,y,t) \right\rangle =&  \left(\int\limits_{0}^{1}\mathrm{d} y\hat{T}_{0} (0,y) \right)^{3}
\displaystyle  \frac{1}{ \left( 2\pi t \right)^{\frac{3}{2}}\sqrt{(\lambda^{(2)} -\lambda^{(1,1)})^{2}(\lambda^{(2)}+2\lambda^{(1,1)})}}+\mathcal{O} (t^{- \frac{5}{2}}). \\
\end{array}
\end{equation}
Here $\lambda^{(2)},\lambda^{(1,1)}$ can be obtained by the perturbation method introduced in the appendix of \cite{bronski1997scalar}. When $\xi(t)$ is a Gaussian white noise process, the derivatives of eigenvalues in the \eqref{eq:firstThreeMoments} are
\begin{equation}\label{eq:eigenvalueWhite}
\begin{array}{rl}
& \lambda^{(2)}=2+ \text{Pe}^2\int\limits_{}^{1}\mathrm{d}y\, u^2(y)\\
& \lambda^{(1,1)}= \text{Pe}^{2}\left( \int\limits_{0}^{1}\mathrm{d} y \,u(y) \right)^{2}=\text{Pe}^{2}\bar{u}^{2}\\
\end{array}
\end{equation}

Conversely, when $\xi(t)$ is the stationary Ornstein-Uhlenbeck process,  the derivatives of eigenvalues in the \eqref{eq:firstThreeMoments} are
\begin{eqnarray}\label{eq:eigenvalueOU}
  &&\hspace{-2cm} \lambda^{(2)}= 2+\text{Pe}^{2} \sqrt{\gamma } \int_{0}^{1}\mathrm{d}y \,u(y) \left\{\frac{\cosh \left(\sqrt{\gamma } y\right)}{\sinh\left(\sqrt{\gamma }\right)} 
  \int_{0}^{1}\mathrm{d}s \,u(s) \cosh \left(\sqrt{\gamma } \left(1-s\right)\right) \right.
 \nonumber   
\\
&&\hspace{5cm}\left.
-\int_{0}^y \mathrm{d}s \,u(s) \sinh \left(\sqrt{\gamma } (y-s)\right) \right\}   
  \\
 &&\hspace{-2cm}
 \lambda^{(1,1)}= \text{Pe}^{2}\bar{u}^{2}
   \nonumber
\end{eqnarray}
The white noise can be regraded as a limiting case of vanishing correlation time $\gamma^{-1}$ in the stationary Ornstein-Uhlenbeck process. It is natural to ask whether the scalar field statistics with $\xi(t)$ an Ornstein-Uhlenbeck process asymptotically satisfies, as $\gamma\rightarrow \infty$, the corresponding model with  white noise process. In the free-space problem, Resnick \cite{resnick1996dynamical} proves this for linear shear flow $u(y)=y$ via the exact formula of $\Psi_{N}$. In channel domain problem, the asymptotic analysis shows that equation \eqref{eq:eigenvalueOU} converges to equation \eqref{eq:eigenvalueWhite} as $\gamma\rightarrow+\infty$, which supports this compatibility for large values of the parameter $\gamma$. In the free space problem, \cite{vanden2001non} proves that both  of the two flows we considered in this paper share the same limiting distribution of the scalar field at long times for any $\gamma$. However, in channel domains, the differences between equations \eqref{eq:eigenvalueWhite} and \eqref{eq:eigenvalueOU} lead to different corresponding limiting distributions. Thus, impermeable boundaries can affect the limiting distribution of the random scalar fields.

The right hand side of each equation in \eqref{eq:NpointCorrelation} is independent of $x,y$, which means all points in the domain have the same statistics behavior at long times. Without loss of generality, we focus on the single point $T(0,0,t)$ of the random scalar field.
 In \cite{camassa2019symmetry}, the authors derived the PDF of $T(0,0,t)$ at long times for the free space version of \eqref{eq:advectionDiffusionNonDimension} and $u(y)=y$, $\text{Pe}=1$ using the method of characteristics and  the Green function. The study of the explicit formula of PDF for the free space problem shows that the skewness of the PDF for $T(0,0,t)$ is positive at long times while the numerical studies show the skewness becomes negative in presence of impermeable channel boundaries, demonstrating how the impermeable boundary has a crucial impact on the PDF of random scalar flied. With the long time asymptotic expansion of moments \eqref{eq:firstThreeMoments} at hand, we can theoretically study the skewness of $T(0,0,t)$ for various parameters and more general shear flows.

\begin{figure}
   \centering
     \includegraphics{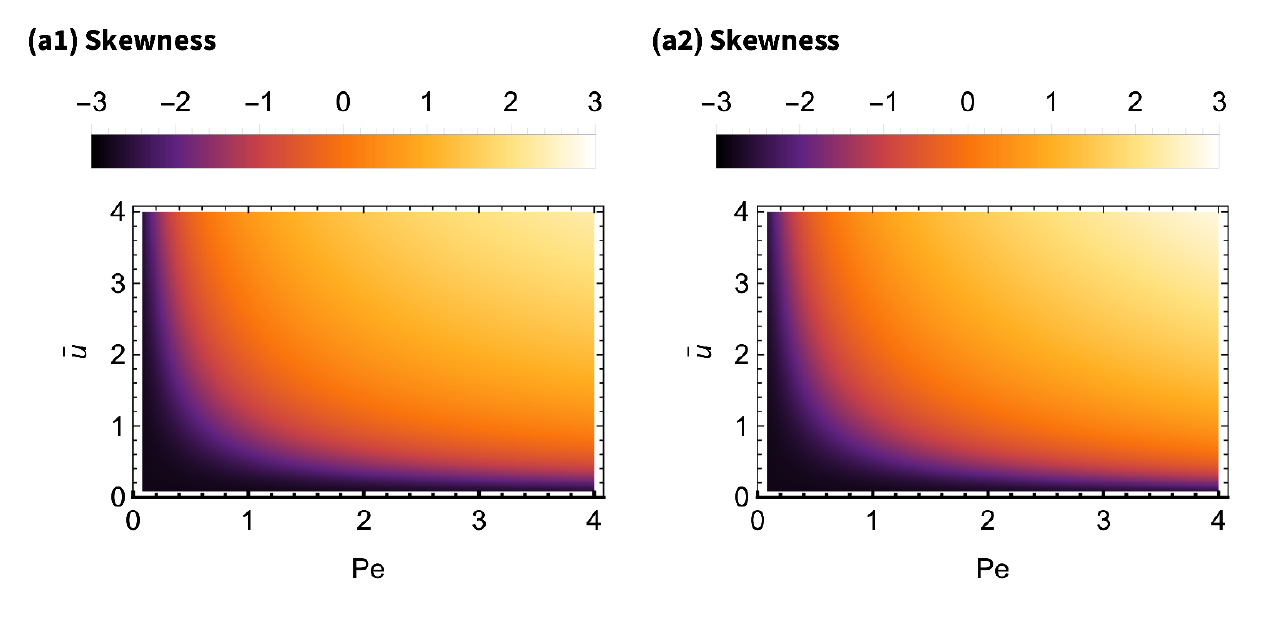}
  \hfill
  \caption[ ]
  {\textbf{The skewness limit of $T(0,0,t)$ at long times for various P\'eclet numbers and $\bar{u}$.} In both panels, the flow takes the form $\text{Pe}\,u(y)\xi(t)$, where $u(y)=(y+A)$ and $\bar{u}=A-\frac{1}{2}$. In panel (a1), $\xi(t)$ is the Gaussian white noise process. In panel (a2), $\xi(t)$ is the stationary Ornstein-Uhlenbeck process with $\gamma=1$. }
  \label{fig:SkewnessShearPeA}
\end{figure}

\begin{figure}
   \centering
     \includegraphics{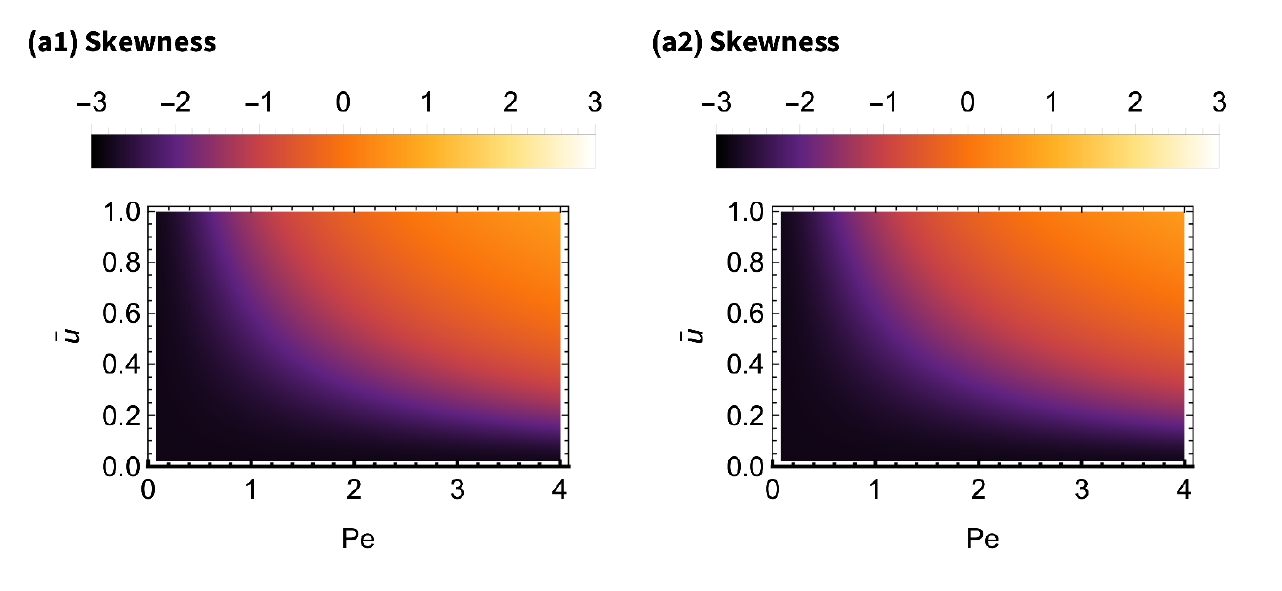}
  \hfill
  \caption[ ]
  {\textbf{The skewness limit of $T(0,0,t)$ at long times for various P\'eclet numbers and $\bar{u}$.} In both panels, the flow takes the form $\text{Pe}u(y)\xi(t)$, where $u(y)=\theta (a-y)$ and  $\bar{u}=a$. In panel (a1), $\xi(t)$ is the Gaussian white noise process. In panel (a2), $\xi(t)$ is the stationary Ornstein-Uhlenbeck process with $\gamma=1$. }
  \label{fig:SkewStepPea}
\end{figure}

\begin{figure}
   \centering
     \includegraphics{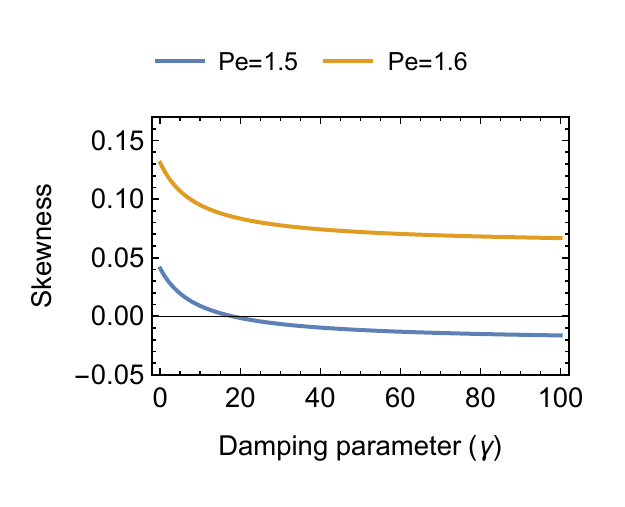}
  \hfill
  \caption[ ]
  {\textbf{The skewness limit of $T(0,0,t)$ at long times for various damping parameters $\gamma$.} The flow is $\text{Pe}(y+1.2)\xi(t)$, where $\xi (t)$ is a stationary Ornstein-Uhlenbeck process with the damping parameter $\gamma$. The cases of $\text{Pe}=1.5$ and $\text{Pe}=1.6$ are shown by the blue and orange curves, respectively.}
  \label{fig:skewshearGamma}
\end{figure}

Based on the formula in \eqref{eq:firstThreeMoments}, as $t\rightarrow \infty$, the variance of $T(x,y,t)$ is given by
\begin{equation}\label{eq:variaceLongTimeAsymptotic}
\begin{array}{rl}
\text{Var}(T)= & \left\langle (T- \left\langle T \right\rangle)^{2} \right\rangle\\
 = & \left( \displaystyle\int_{0}^{1}\mathrm{d} y\hat{T}_{0} (0,y) \right)^{2}  \left( \displaystyle\frac{1}{ \sqrt{(\lambda^{(2)})^{2}-(\lambda^{(1,1)})^{2} }} -\frac{1}{\lambda^{(2)}} \right)  \displaystyle \frac{1}{2\pi t} +\mathcal{O}(t^{-2}) .
 \\
\end{array}
\end{equation}
Notice that coefficient of $t^{-1}$ in \eqref{eq:variaceLongTimeAsymptotic} is strictly positive if $\lambda^{(1,1)}\neq 0$, which requires $\bar{u} \neq 0$. As $t\rightarrow \infty$, the skewness of $T(x,y,t)$ is given by
\begin{equation}\label{eq:skewnessLongTimeAsymptotic}
\begin{array}{rl}
\text{S}(T)=& \displaystyle \frac{ \left\langle (T- \left\langle T \right\rangle)^{3} \right\rangle}{\left( \hbox{\small Var}(T) \right)^{\frac{3}{2}}}\\\\
=&\displaystyle \frac{
 \displaystyle\frac{1}{\sqrt{(\lambda^{(2)} -\lambda^{(1,1)})^{2}(\lambda^{(2)}+2\lambda^{(1,1)})}}
- \displaystyle\frac{3}{\sqrt{\lambda^{(2)} (\lambda^{(2)})^{2}-(\lambda^{(1,1)})^{2} }}
+ \displaystyle\frac{2}{(\lambda^{(2)})^{\frac{3}{2}}}
}{\left( \displaystyle \frac{1}{ \sqrt{(\lambda^{(2)})^{2}-(\lambda^{(1,1)})^{2} }} - \displaystyle\frac{1}{\lambda^{(2)}}  \right)^{\frac{3}{2}}}+\mathcal{O}(t^{-1}).\\
\end{array}
\end{equation}
The first term on the right hand side of \eqref{eq:skewnessLongTimeAsymptotic} shows the existence of tbe long time limit of skewness, which means the PDF of $T(0,0,t)$ has a persisting asymmetry. There are five factors that affect the limit value: the P\'eclet number $\text{Pe}$, the  mean of spatial component of flow $\bar{u}$, the shape of $u(y)$, the temporal fluctuation $\xi(t)$ and the OU damping parameter  $\gamma$.  Figure \ref{fig:SkewnessShearPeA} and figure \ref{fig:SkewStepPea} show the long time limit of skewness of $T(0,0,t)$ for the flow with $u(y)=y+A$ and
$u(y)=\theta(a-y)$ (with $\theta$ denoting the Heaviside step function)
for various P\'eclet numbers and $\bar{u}$, respectively.  Panel  (a1) in figure \ref{fig:SkewnessShearPeA} shows that the skewness limit is negative when $\bar{u}={1}/{2}$, ${Pe}=1$, which is consistent with Monte-Carlo simulation results reported in~\cite{camassa2019symmetry}. Both of those figures have a similar pattern. The skewness is negative when \text{Pe} or $\bar{u}$ is small and positive when they are large.  Alternatively, with the step function shear flow, the differences are larger as can be seen by comparing panel (a1) or  panel (a2) in figure \ref{fig:SkewnessShearPeA} with the corresponding panels in figure \ref{fig:SkewStepPea}. One can see that the change of $u(y)$ dramatically changes the long time asymptotics of the skewness in its $\text{Pe}-\bar{u}$ dependence.  Of course, while the Ornstein-Uhlenbeck process yields different numerical values compared with the white noise process, for the parameter region $\text{Pe}\times \bar{u} \in [0,4]\times [0,4]$ shown in the figures the relative difference between them is less than $0.1$. Hence it is hard to observe a difference when comparing the left panel and the right panel in figure \ref{fig:SkewnessShearPeA} or figure \ref{fig:SkewStepPea}.  Figure \ref{fig:skewshearGamma} shows the dependence of the skewness long time limit on the damping parameter $\gamma$. Note that depending on $\text{Pe}$ and $u(y)$, the sign of skewness can be made to change by varying $\gamma$.

\section{An explicit example for scalar intermittency}
\label{sec:wind}
 In this section we study a special case of \eqref{eq:advectionDiffusion}, which yields an exact formula valid at all times. Therefore, this is a solid benchmark for the long time asymptotic analysis derived in the previous section. In \cite{bronski2007explicit}, the authors call the advection-diffusion equation \eqref{eq:advectionDiffusion} with $u(y)=1$   the `wind model'. They  study the one dimensional problem when $\xi(t)$ is the Gaussian white noise process. Here, we present the exact formula of $N$-point correlation function $\mathbf{\Psi}_N$ for the channel domain problem with any general Gaussian process $\xi(t)$.

The associated Green's function $G(x,y,x_{0},y_{0},t)$, that is, the solution of \eqref{eq:advectionDiffusion} with the initial condition $T(x,y,0)=\delta(x-x_{0})\delta(y-y_{0})$, can be obtained by the separation of variables and the method of characteristics,
\begin{equation}
\begin{array}{rl}
G(x,y,x_{0},y_{0},t)=&K(y,y_{0},t) \displaystyle\frac{1}{ \sqrt{4 \pi t}}\exp \Bigg( \displaystyle - \frac{(x-x_{0}-\text{Pe} \int\limits_0^t\mathrm{d}s \,\xi(s) )^{2}}{4t} \Bigg) .\\
\end{array}
\end{equation}
where $K(y,y_{0},t)= 1+ 2 \sum\limits_{n=1}^{\infty} \cos (n\pi y) \cos (n \pi y_0) \exp (-n^2 \pi^2 t)$. The solution with general initial condition $T_0(x,y)$ can be constructed by the Green function via convolution,
\begin{equation}
\begin{array}{rl}
T(x,y,t)= & \int\limits_{-\infty}^{+\infty} \mathrm{d}x_0 \int\limits_{0}^{1}  \mathrm{d} y_{0} T_0 (x_{0},y_0) G(x,y,x_{0},y_{0},t). \\
\end{array}
\end{equation}
By the definition of $\mathbf{\Psi}_N$ and Fourier transform, we have
\begin{equation}
\begin{array}{rl}
&\mathbf{\Psi}_N (\mathbf{x}, \mathbf{y},t)=\\
 & \int\limits_{\mathbb{R}^{N}}^{}\mathrm{d}\mathbf{x}_0 \int\limits_{[0, 1]^{N}}^{} \mathrm{d} \mathbf{y}_{0} \frac{1}{(2\pi)^N}\int\limits_{\mathbb{R}^{N}}^{} \mathrm{d} \mathbf{k}\text{exp} ( \sum\limits_{j=1}^N -t  k_j^2-\mathrm{i}k_j (x_{j}-x_{0j}))\left\langle \text{exp} (\mathrm{i} \text{Pe} \int\limits_0^t\mathrm{d}s \xi(s) \sum\limits_{j=1}^Nk_j) \right\rangle\\
&\times \prod\limits_{j=1}^NK(y_{j},y_{0j},t )T_{0}(x_{0j},y_{0j}) ,\\
\end{array}
\end{equation}
where $\mathbf{x}_{0}=\left(x_{01},x_{02},\cdots,x_{0N}\right)$, $\mathbf{y}_{0}=(y_{01},y_{02},\cdots,y_{0N})$. Since $\xi(t)$ is a Gaussian process, $\int\limits_0^t\mathrm{d} s \xi(s)$ is a normal random variable at any instant of time. By utilizing the characteristic function of the normal random variable, we obtain:
\begin{equation}
\begin{array}{rl}
&\mathbf{\Psi}_N (\mathbf{x}, \mathbf{y},t)=\\
 &\int\limits_{\mathbb{R}^{N}}^{}\mathrm{d}\mathbf{x}_0 \int\limits_{[0, 1]^{N}}^{} \mathrm{d} \mathbf{y}_{0} \frac{1}{(2\pi)^N}\int\limits_{\mathbb{R}^{N}}^{}\mathrm{d} \mathbf{k}\, \exp ( \sum\limits_{j=1}^N -t  k_j^2-\mathrm{i}k_j (x_{j}-x_{0j})) \text{exp}\left( - \frac{1}{2} v(t)  (\text{Pe}\sum\limits_{j=1}^{N}k_j)^{2} \right)\\
&\times \prod\limits_{j=1}^NK(y_{j},y_{0j},t )T_{0}(x_{0j},y_{0j})\\
\end{array}
\end{equation}
where $v(t)$ is the variance of stochastic process $\int\limits_0^t\mathrm{d} s \xi(s)$. Comparing this integral with the multivariate normal distribution, we have
\begin{equation}
\begin{array}{rl}
\mathbf{\Psi}_N (\mathbf{x}, \mathbf{y},t)= &\int\limits_{\mathbb{R}^{N}}^{}\mathrm{d}\mathbf{x}_0 \int\limits_{[0, 1]^{N}}^{} \mathrm{d} \mathbf{y}_{0}  \frac{\text{exp}\left( -\frac{1}{2}(\mathbf{x}-\mathbf{x}_{0})\Lambda^{-1}(\mathbf{x}-\mathbf{x}_{0})^{\text{T}} \right)}{(2\pi)^\frac{N}{2}\sqrt{\text{det} (\Lambda)} }\prod\limits_{j=1}^NK(y_{j},y_{0j},t )T_{0}(x_{0j},y_{0j})\\
\end{array}
\end{equation}
where $\Lambda= 2tI+v(t) \text{Pe}^2\mathbf{e}^{\text{T}}\mathbf{e}$, $I$ is the identity matrix of size $N\times N$ and $\mathbf{e}$ is a $1\times N$ vector with $1$ in all coordinate. By the Sherman-Morrison formula \cite{sherman1950adjustment}, $\Lambda^{-1}=(2t)^{-1}\left( I- \frac{v(t)\text{Pe}^2\mathbf{e}^{\text{T}}\mathbf{e} }{2t+ N v(t)\text{Pe}^2} \right)$, and by the matrix determinant lemma, $\text{det} (\Lambda)= (2t)^{N}\left( 1+ \frac{N v(t)\text{Pe}^2}{2t} \right)$.
To compare with the $N$-th moment $\left\langle T^N(x,y,t) \right\rangle$, we choose the initial condition as $T_0(x,y)=\delta(x)$. Hence, the solution is independent of $x$, 
\begin{equation}
\begin{array}{rl}\label{eq:LinesourceWind}
T(x,t)=\displaystyle\frac{1}{ \sqrt{4 \pi t}}\text{exp} \Bigg( - \frac{(x-\text{Pe} \int\limits_0^t\mathrm{d}s \xi(s) )^{2}}{4t} \Bigg),
\end{array}
\end{equation}
and the $N$-th moment is
\begin{equation}
\begin{array}{rl}\label{eq:NmomentWind.pdf}
\left\langle T^{N}(x,t) \right\rangle=& \displaystyle\frac{1}{(4\pi t)^{\frac{N}{2}}} \frac{1}{\sqrt{1+\frac{N v(t)\text{Pe}^2}{2t} }}\text{exp}\left( -\frac{N x^{2}}{4t}\left(1-\frac{N \text{Pe}^2 v (t)}{N \text{Pe}^2 v (t) +2 t}\right) \right). \\
\end{array}
\end{equation}

\begin{figure}
 \centering
 \includegraphics{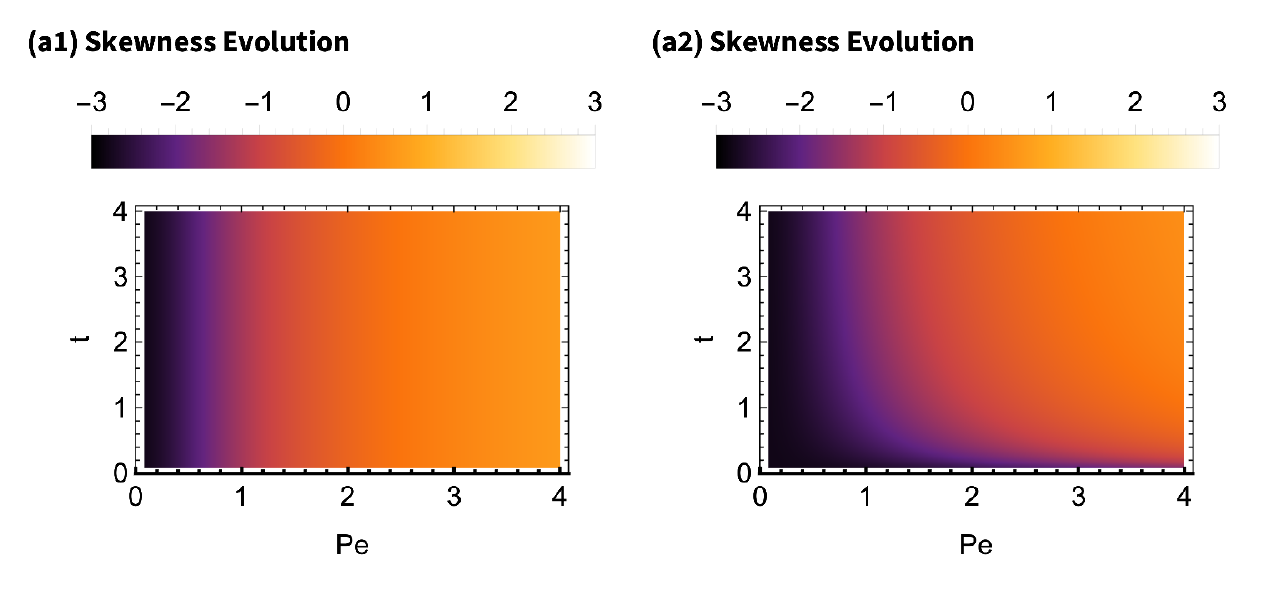}
 \hfill
 \caption[ ]
 {\textbf{Evolution of the random variable $T(0,0,t)$ for various P\'eclet numbers and time}. In both panels, the flow takes the form $\text{Pe}u(y)\xi(t)$, where $u(y)= 1$ and  $\bar{u}=1$. In panel (a1), $\xi(t)$ is the Gaussian white noise process. In panel (a2), $\xi(t)$ is the stationary Ornstein-Uhlenbeck process with $\gamma=1$. }
 \label{fig:SkewnessPet}
\end{figure}

\begin{figure}
 	\centering
 	\includegraphics{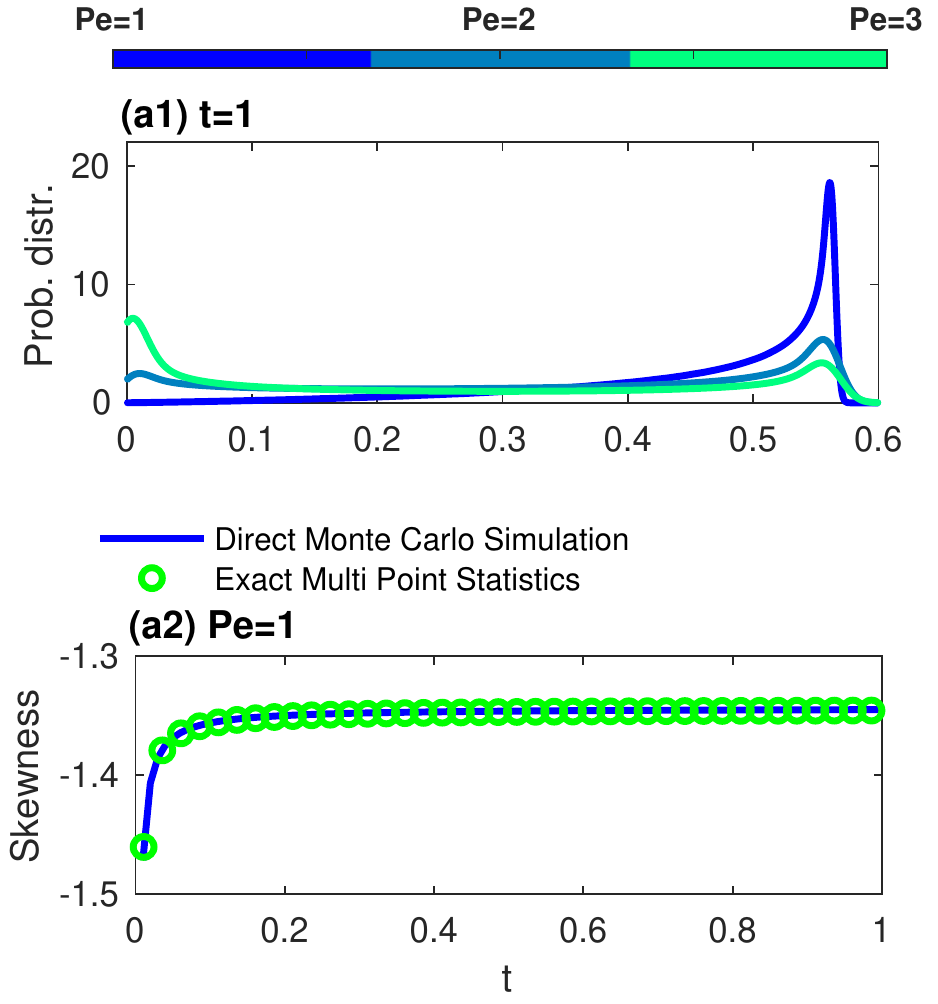}
 	\caption{\textbf{Evolution of the spatially averaged random variable defined in \eqref{eq:WindStrip} for various P\'eclet numbers and time}.  In panel (a1), we superposed $3$ probability distributions of $M (\frac{1}{10},1)$ with $\text{Pe}=1$ (blue), $\text{Pe}=2$ (dark green) and $\text{Pe}=3$ (green). Next, in panel (a2), we show the skewness of $M (\frac{1}{10},t)$ calculated through the Direct Monte Carlo simulations (blue line) and the numerically integrating the formula \eqref{eq:NmomentWindStrip} (green circle). }
 	\label{fig:wind_strength}
 \end{figure}

\begin{figure}
 \centering
 \includegraphics{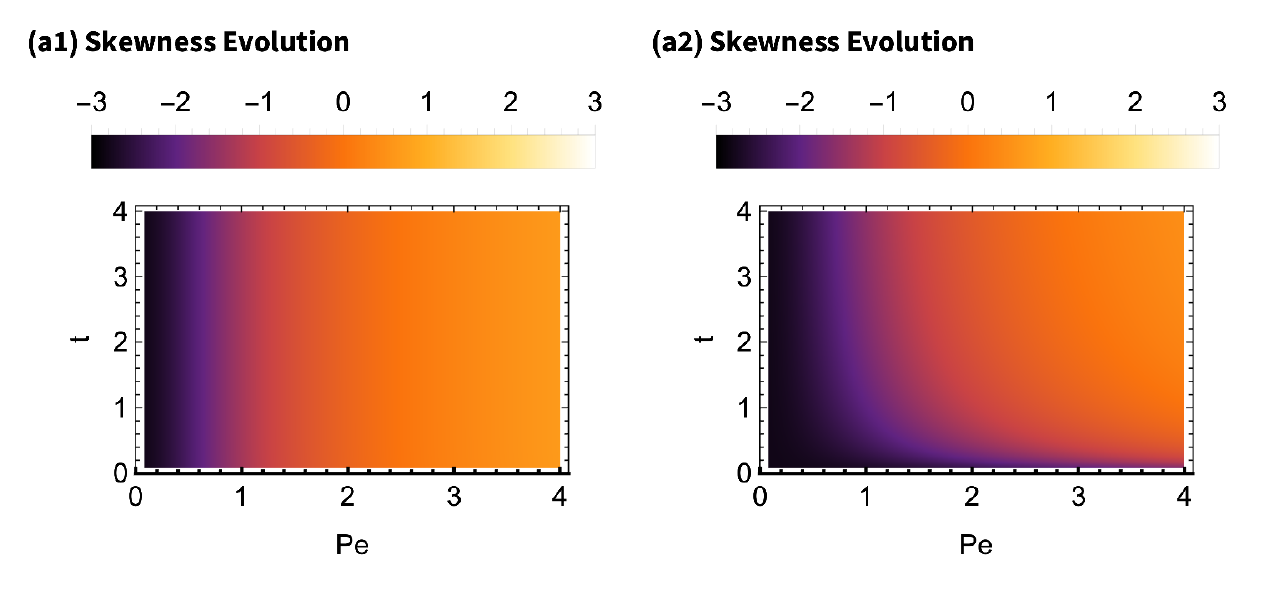}
 \hfill
 \caption{
{ \textbf{Evolution of the spatially averaged random variable $M( \frac{1}{10},t)$ defined in \eqref{eq:WindStrip} for various P\'eclet numbers and time}. In both panels, the flow takes the form $\text{Pe}u(y)\xi(t)$, where $u(y)= 1$ and  $\bar{u}=1$. In panel (a1), $\xi(t)$ is the Gaussian white noise process. In panel (a2), $\xi(t)$ is the stationary Ornstein-Uhlenbeck process with $\gamma=1$. }
 \label{fig:SkewnessStripPet}}
\end{figure}

The long time asymptotic expansion of \eqref{eq:NmomentWind} is consistent with the  \eqref{eq:firstThreeMoments}.  Figure \ref{fig:SkewnessPet} shows the skewness evolution of the scalar field at the point $(0,0)$ varying with different $\text{Pe}$.  
To consider a more multi-point statistic as studied in the Monte-Carlo simulations in \cite{camassa2019symmetry}
we may study the average of the scalar field over $x\in[-a,a]$,
\begin{equation}\label{eq:WindStrip}
\begin{array}{rl}
M(a,t)= & \displaystyle\frac{1}{2a}\int\limits_{-a}^aT(x,t)\mathrm{d}x= \frac{1}{2a} \left(\text{erf}\Bigg(\frac{a+\text{Pe} \int\limits_0^{t}\mathrm{d} s\xi(s)}{2 \sqrt{t}}\Bigg)+\text{erf}\Bigg(\frac{a-\text{Pe} \int\limits_0^{t}\mathrm{d} s\xi(s)}{2 \sqrt{t}}\Bigg)\right),\\
\end{array}
\end{equation}
where $\text{erf}(z)= \frac{2}{\sqrt{\pi}} \int\limits_0^{z}\mathrm{d} t e^{-t^2}$ is the error function. When $a\rightarrow 0$, $M(a,t)$ will converge to $T(0,t)$.
By switching the order of integration and ensemble average, the $N$-moment of $M(a,t)$ is
\begin{equation}\label{eq:NmomentWindStrip}
\begin{array}{rl}
\left\langle M(a,t)^N \right\rangle= & \frac{1}{a^{N}}\int\limits_{[-a,a]^N}^{}\mathrm{d} \mathbf{x}\left\langle  \prod\limits_{j=1}^N T(x_{j},t) \right\rangle
\end{array}
\end{equation}
To verify our theoretical analysis, we perform the Direct Monte-Carlo(DMC) method proposed in \cite{camassa2019symmetry}. Panel (a2) in  figure \ref{fig:wind_strength} shows the skewness computed by the theoretical approach and DMC approach. The consistency of the two approaches demonstrates the validity of the theoretical analysis in this section. Panel (a1) in  figure \ref{fig:wind_strength} depicts the PDF of $M(\frac{1}{10},1)$ obtained by DMC method for different P\'{e}clet numbers. It shows that with increasing P\'{e}clet number the PDF changes  from  negatively-skewed to positively-skewed, which is consistent with the observation we made from figure \ref{fig:SkewnessShearPeA} and \ref{fig:SkewStepPea}.  Figure \ref{fig:SkewnessStripPet} shows the skewness evolution of $M(\frac{1}{10},t)$ computed by \eqref{eq:NmomentWindStrip} for various P\'eclet numbers and different types of temporal fluctuation. Panel (a1) in figure \ref{fig:SkewnessStripPet} shows the skewness evolution when $\xi (t)$ is the white noise process. The skewness is almost unchanged in time, which means that the system reaches the long time asymptotic state in very short time.  Panel (a2) in the same figure shows the skewness evolution when $\xi (t)$ is the stationary Ornstein-Uhlenbeck process. The finite correlation time in the temporal fluctuation $\xi(t)$ introduces a noticeable transient before reaching the long time asymptotic state. This phenomenon weakens as the P\'eclet number increases.

\section{Numerical simulations}
\label{sec:numerical}
To verify the long-time asymptotic analysis results, we need to simulate the sample of random scalar field $T(x,y,t)$ at a single point $(x,y)$. The Feynman-Kac's based backward Monte-Carlo method is efficient in this case, since it can access the single point value of the scalar field without the global solution of \eqref{eq:advectionDiffusion}. For each realization of the stochastic process $\xi(t)$, the random field has the path integral representation $T(x,y,t)=  \left\langle T_0(X_{t} (t),Y_{t} (t)) \right\rangle_{B_{1} (t),B_{2} (t)}$ by the Feynman-Kac's formula,
where $X_{t} (s),Y_{t} (s)$ are the solution of the stochastic differential equation(SDE)
\begin{equation}
\begin{array}{rl}
\mathrm{d}X_{t} (s)&=-\text{Pe}\xi (t-s)u(Y_{t} (s))\mathrm{d}s+ \sqrt{2} \mathrm{d}B_{1} (t)\\
\mathrm{d}Y_{t} (s)&=\sqrt{2}\mathrm{d}B_{2} (t) \\
X_{t} (0)=x& \quad Y_{t}(0)=y\\
\end{array}
\end{equation}
where  $B_i$ are independent Brownian motions. Notice that both the white noise process and the stationary Ornstein-Uhlenbeck process are stationary and temporally homogeneous, so $\xi(t-s)=\xi(s)$. This property allow us to reuse the solution$(X_{t_{i}} (s),Y_{t_{i}} (s))$ to compute $(X_{t_{i+1}} (s),Y_{t_{i+1}} (s))$, which saves a lot of computation cost. We solve the SDE by an Euler scheme with a time increment $\Delta s =0.01$.
\begin{equation}
\begin{array}{rl}
X_{s_{i+1}}&=X_{s_{i}}-\text{Pe}\xi (s_{i})u(Y_{t} (s_{i}))\Delta s++ \sqrt{2 \Delta s}n_{1,i}\\
Y_{s_{i+1}}&=Y_{s_{i}}+\sqrt{2\Delta s}n_{2,i}  \\
\end{array}
\end{equation}
$n_{1,i},n_{2,i}$ are standard independent and identically distributed normal random variables which are produced by the Mersenne Twister uniform random number generator. We impose billiard-like reflection rules on the boundary $y=0,1$.  We typically generate  $10^{6}$ realization of $\xi(s)$. The realization of the Gaussian white noise process are produced by $\xi(t_{i})= \frac{n}{\sqrt{\Delta s}}$. The Ornstein-Uhlenbeck process are simulated by the scheme in \cite{gillespie1996exact}. For each realization of $\xi (s)$, we use $10^6$ independent SDE solution $(X_{t}(s), Y_{t} (s))$ to compute the path integral representation of $T(x,y,t)$. The simulations are performed on UNC's Longleaf computing cluster with 400 parallel computing jobs, and each job takes approximately 3 days on the cluster.

\begin{figure}[tbp]
  \centering
    \includegraphics[width=1\linewidth]{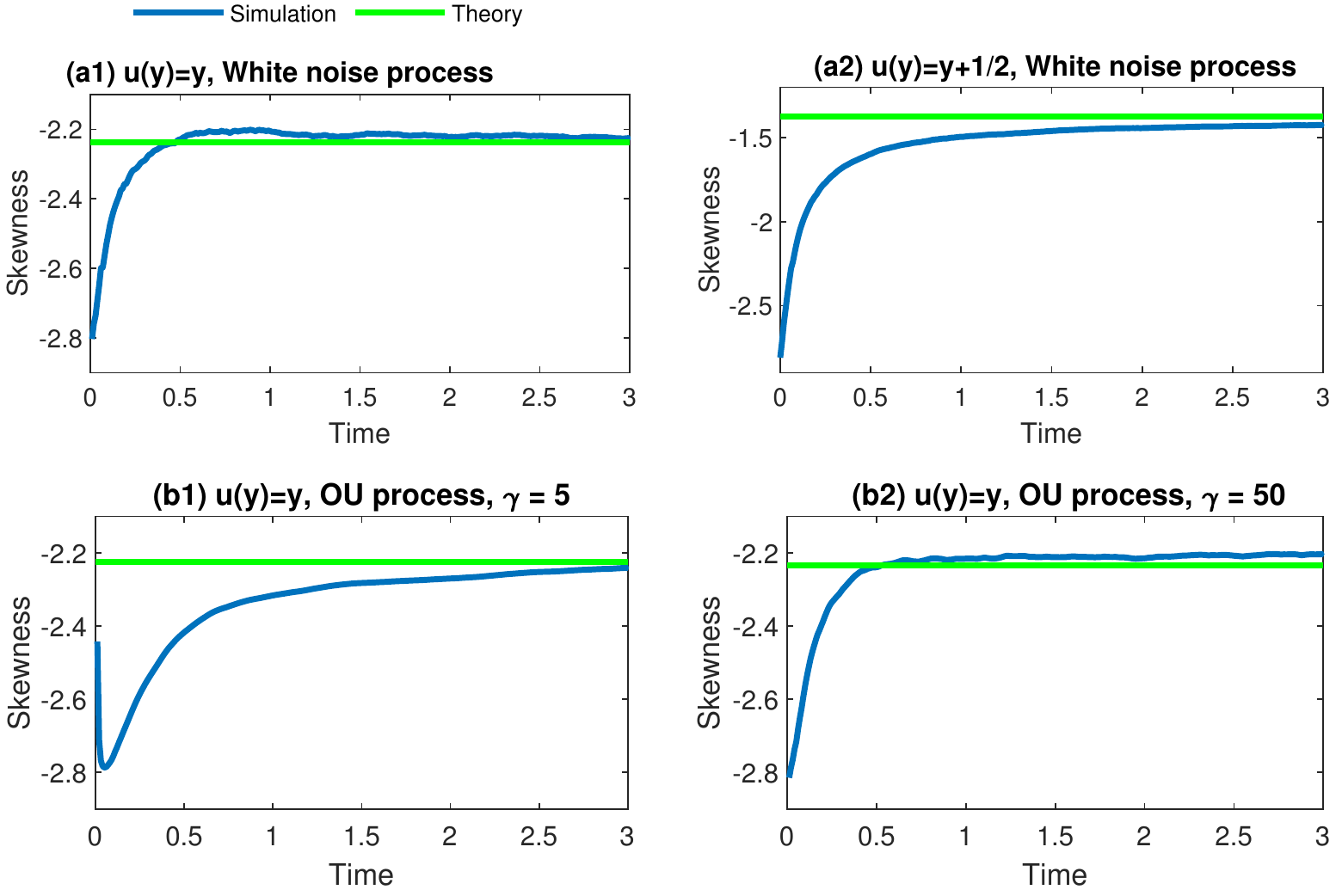}
  \hfill
  \caption{\textbf{Skewness evolution for random shear flows fluctuating with Gaussian white noise and Ornstein-Uhlenbeck process statistics.}
  {Here, we provide the skewness evolution for random shear flows with different spatial means and different fluctuation statistics obtained from Monte Carlo simulations along with the long time asymptotics theoretical predictions of skewness \eqref{eq:skewnessLongTimeAsymptotic}. In panels (a1)-(a2) we provide the skewness evolution and corresponding long time asymptotics for flows with Gaussian white noise fluctuations. Next, in panels (b1)-(b2), we provide the skewness evolution and its long time asymptotics for random shear flow $u(y)=y$ with fluctuations that has Ornstein-Uhlenbeck process statistics. Furthermore, in panel (b1) and (b2), the correlation strength parameter of Ornstein-Uhlenbeck processes are $\gamma=5$ and $\gamma=50$, respectively.
  In all panels, Monte Carlo simulation results are shown in blue curves where the theoretical predictions of the long time limits of skewness are shown in green horizontal line.}
  \label{fig:SkewnessTimeFK}}
\end{figure}

We simulate $T(0,0,t)$ with the initial condition $T_0(x,y)= {e^{-x^2}}/{\sqrt{\pi}}$ for different flows, and results are shown in figure \ref{fig:SkewnessTimeFK}. The blue curves are the numerical result of skewness evolution and the green horizontal lines are the skewness limits computed by \eqref{eq:skewnessLongTimeAsymptotic}. The consistency between them validates this formula.
In panel (a1) and (a2) , $\xi(t)$ is white noise process,  $u(y)$ are $y$ and $y+\frac{1}{2}$ respectively. One can see that the larger spatial mean of the flow leads to longer transient dynamics before reaching the long time asymptotic state.  In panel (b1) and (b2) , $\xi(t)$ is the stationary Ornstein-Uhlenbeck process and  $u(y)=y$, the damping parameter $\gamma$ are $5$ and $50$ respectively. Comparison between  panel (b1) and (b2) shows the longer correlation time in the panel (b1) yields a more dramatic transient dynamics in the skewness evolution.
Comparing the panel $(a1)$ and panel $(b2)$, we see the convergence of the Ornstein-Uhlenbeck case to the white noise case when the correlation time $\gamma^{-1}$ is small.


\section{Conclusion}
\label{sec:discuss}
We have demonstrated analytically and numerically that the single point statistics, in particular \emph{skewness}, of a passive scalar advected by a random shear flow with deterministic initial data have opposite symmetric behaviors at long times depending on the presence or the absence of  \emph{impermeable boundaries} . We have investigated two types of flow temporal fluctuations, respectively modeled by Gaussian white noise and stationary Ornstein-Uhlenbeck processes. We have shown the convergence of the Ornstein-Uhlenbeck case to its white noise counterpart in the limit  
$\gamma\rightarrow \infty$ of the OU damping parameter, which generalizes the conclusion in the article \cite{resnick1996dynamical}  for free space  to the confined channel domain problem.  Importantly, we observe that the OU damping parameter $\gamma$ plays a more significant role in channel domains than in the free space problem. The first three moments of the scalar distribution at infinite time depend on the correlation time $\gamma^{-1}$ in the channel domain, which is in strong contrast to the result of Vanden-Eijnden \cite{vanden2001non} in free space where the PDF at long time is independent of $\gamma$.    We have presented the detailed discussions of three different shear flows. All of them show the transient of skewness from negative to positive when increasing either the P\'eclet number or $\bar{u}$, which rigorizes and generalizes the observation from the simulation result in \cite{camassa2019symmetry}. To find a benchmark for theoretical analysis, we have generalized the wind model studied in \cite{bronski2007explicit} and derived the exact formula of the $N$-point correlation function for the flow with no spatial dependence and Gaussian temporal fluctuation. The long time asymptotic expansion of this formula is consistent with our theory for general shear flows.

We have presented numerical studies that verify the validity of our theoretical results.  We have performed Direct Monte Carlo simulation for the wind model and observed that the P\'eclet number can adjust the time at which the skewness of the distribution changes sign. Due to the lack of an exact solution for general shear flows, we implemented  backward Monte-Carlo simulations to verify the long time asymptotic results we derived. We confirmed that as the damping parameter $\gamma$ increases the stationary Ornstein-Uhlenbeck case converges to the white noise case and found that transient for the skewness of the passive scalar's PDF to reach its long time asymptotic state lasts longer as the damping parameter decreases. 

Future work will include considering an experimental campaign with the associated theoretical analysis. Our recent study \cite{ding2020enhanced} regarding the enhanced diffusion \cite{taylor1953dispersion} and third spatial Aris moment \cite{aris1956dispersion} induced by a periodically moving wall led to the development of an experimental framework of the model explored in this paper.  The computer controlled robotic arm we developed for the periodic study can be applied to the case of a randomly moving wall, such as the OU process $\xi(t)$, with suitable parameters for the fluid and the channel. The induced flow in the channel can be modeled by $y\xi(t)$. Hence, the tracers in the fluid satisfy the advection-diffusion equation \eqref{eq:advectionDiffusion}. The symmetry properties of the tracer's PDF can be predicted by the theory we developed here. Perhaps even more interesting will be considering cases in which the physical shear flow is not decomposed into a product of a function of space and a function of time, such as happens with the general nonlinear Ferry wave solutions at finite viscosities.  More involved analysis will clearly be needed to study these interesting configurations.

Lastly, we discuss and overview interesting issues associated with a vanishing spatial mean of the flow.  Such a case could be experimentally observed by having two walls executing equal but opposite parallel motions, or by putting the observer in an appropriate frame of reference.  The asymptotic analysis strategy we presented doesn't technically fail if the spatial average of the flow in the physical domain vanishes, namely $\bar{u}= \int_{0}^{1}\mathrm{d} y u(y) =0$, as the distribution is expected to be symmetric at long time in this case, consistent with our asymptotics which show that the third moment vanishes at long time with a zero spatial mean.  However, the case with $\bar{u}=0$ requires considerable additional analysis to investigate how the long time PDF relaxes to a symmetric state. To see that, it is easy to check that the coefficient of $t^{-1}$ in the equation \eqref{eq:variaceLongTimeAsymptotic} becomes zero as well as the coefficient of $t^{-\frac{3}{2}}$ and $t^{- \frac{5}{2}}$ in centered third moment expansion. That means the higher-order terms in \eqref{eq:firstThreeMoments} are needed for the analysis of skewness for the case $\bar{u}=0$. The thesis work of \cite{Kilic} reports preliminary results that the point statistics induced by some flows with $\bar{u}=0$ have distinct behaviors from the case $\bar{u}\neq 0$. More detailed analysis in this direction has been completed and it will be reported separately.

\section{Acknowledgements}

We acknowledge funding received from the following NSF Grant Nos.: DMS-1517879, and DMS-1910824; and ONR Grant No: ONR N00014-18-1-2490.

\bibliographystyle{elsarticle-harv}

%
%


\clearpage

\appendix
\numberwithin{equation}{section}
\numberwithin{figure}{section}
\numberwithin{table}{section}


\end{document}